\documentclass[twocolumn]{openjournal}
\pdfoutput=1
\usepackage{graphicx}
\usepackage{enumerate}
\usepackage{amssymb, amsmath, amsfonts}
\usepackage{hyperref}
\usepackage{ulem}
\usepackage{bm}
\usepackage{xspace}
\usepackage{xcolor}

\defcitealias{Lindegren+Dravins_2021}{LD21}

\usepackage{orcidlink}

\begin{document}

\title{Astrometric Radial Velocities from the Hipparcos-Gaia Catalog of Accelerations and Implications for Astrometric Acceleration Measurements}

\author{\vspace{-30pt}Timothy D.~Brandt\,\orcidlink{0000-0003-2630-8073}$^{1}$}
\affiliation{$^1$Space Telescope Science Institute, 3700 San Martin Drive, Baltimore, MD, 21218, USA}
\email{Corresponding author: tbrandt@stsci.edu}

\begin{abstract}
Astrometry from the Gaia satellite and from the long-term combination of Hipparcos and Gaia are now sensitive to sky-plane accelerations as low as $\approx$1\,m\,s$^{-1}$\,yr$^{-1}$.  This paper quantifies and explores an important caveat: apparent nonlinear motion due to a star's nonzero radial velocity can be indistinguishable from real astrometric acceleration.  This nonlinear motion is parallel to the proper motion, so it can be both quantified and avoided by projecting apparent astrometric accelerations into components parallel and perpendicular to the proper motion.  We illustrate this distinction for a sample of very nearby, fast-moving stars from the Hipparcos-Gaia Catalog of Accelerations (HGCA).  We then generalize the effect of stellar radial velocity and projections of the astrometric acceleration to binary stars in which we observe the acceleration of both components.  Finally, we demonstrate that the proper motion differences in the HGCA are statistically well-behaved even for the nearest and fastest-moving stars, at least in the component perpendicular to the proper motion.  This distinction---between astrometric acceleration parallel and perpendicular to proper motion---could have important consequences for future missions reaching extreme astrometric sensitivities around nearby stars.
\end{abstract}

\maketitle

\section{Introduction} \label{sec:introduction}

The apparent position of stars changes with time due to both the Earth's changing perspective and the stars' motion through space.  The former is known as parallactic motion; the latter is called proper motion.  The field of astrometry consists of measurements of apparent stellar positions and fits to model paths across the sky.  The model paths depend on position, proper motion and parallax---the five standard astrometric parameters \citep{HIP_TYCHO_ESA_1997,Perryman_2012}.  

In the limit of small angular motions the apparent sky path is a linear function of the five standard astrometric parameters.  For stars with large proper motions, due to high space velocities and/or proximity to Earth, deviations from this small angle approximation can become measurable.  This is now the case for a handful of nearby stars thanks to the precision of the Gaia spacecraft \citep{Gaia_General_2016,Gaia_General_2018,GaiaDR3}.  

In the small angle approximation, stars moving at constant velocity through space undergo constant proper motion as seen from Earth.  Deviations from constant proper motion can be due to physical accelerations, e.g.~due to massive companions.  This effect has long been used to discover and weigh faint companions to bright stars \citep{Bessel_1844}.  Changes in proper motion also occur due to the breakdown of the small angle approximation.  These latter effects are often termed ``perspective acceleration.''  Relativistic and light travel time effects are also present, but for nearby stars these effects are typically much less important than perspective acceleration \citep{Butkevich+Lindegren_2014}.

If the three-dimensional position and space velocity of a star are known, its sky path may be modeled without the small angle approximation.  The transverse components of the position and velocity may be readily measured from the sky motion itself, while the distance may be measured via the parallax.  The one component of a star's position in phase space that cannot be measured astrometrically is its line-of-sight velocity, or radial velocity (RV).  If a star's RV is known, e.g.~from Doppler spectroscopy, its space motion may be modeled without the small angle approximation, and deviations from the modeled sky path may be used to infer the presence and tug of unseen companions.  Alternatively, the sky path itself may be used to infer an RV under the assumption that a star is undergoing inertial motion (\citealt{Dravins+Lindegren+Madsen_1999}; \citealt{Lindegren+Dravins_2021}, hereafter \citetalias{Lindegren+Dravins_2021}).  

Because RV is a one-dimensional quantity, it can only produce apparent acceleration in a single direction for a given star.  This has consequences for the measurement of astrometric RVs and the expected statistics when optimizing a sky path over all six phase space parameters rather than just the usual five astrometric parameters.  It also impacts the prospects of inferring small physical accelerations in the presence of imperfect knowledge of a star's RV.  

This paper explores the interplay of astrometric acceleration and RVs in the context of the Hipparcos-Gaia Catalog of Accelerations \citep[HGCA,][]{Brandt_2018, Brandt_2021}.  We build on the work of \cite{Dravins+Lindegren+Madsen_1999} and \citetalias{Lindegren+Dravins_2021}, the latter of whom performed a similar analysis to ours using the Hipparcos re-reduction \citep{vanLeeuwen_2007} more directly.  The HGCA gives a calibrated proper motion from the position difference between Hipparcos \citep{HIP_TYCHO_ESA_1997, vanLeeuwen_2007} and Gaia EDR3 \citep{Gaia_EDR3, Lindegren+Klioner+Hernandez+etal_2021}.  The difference between this long-term proper motion and the EDR3 proper motion can be interpreted as a sensitive measure of acceleration.  Given the $\approx$25 year time baseline between Hipparcos and Gaia, the HGCA is very sensitive to perspective acceleration due to the nonzero RVs of nearby stars.  

We structure the paper as follows.  Section \ref{sec:hgca} summarizes the HGCA, while Section \ref{sec:RVimpact} details the effect of the RV on the perspective acceleration.  Section \ref{sec:results_and_notes} summarizes HGCA-based astrometric RV measurements and residuals for nearby stars.  Section \ref{sec:binaries} further develops the statistics for binaries where both binary components have measured accelerations.  Section \ref{sec:statisticalproperties} explores the statistics of the HGCA proper motion differences in the presence of RV-induced apparent accelerations.  Section \ref{sec:future_impact} discusses the impact that uncertain RVs might have on future, highly-sensitive astrometric missions.  We conclude with Section \ref{sec:conclusions}.

\section{The Hipparcos-Gaia Catalog of Accelerations} \label{sec:hgca}

The HGCA consists of three proper motions placed, as closely as possible, into the reference frame of Gaia EDR3: one from Gaia EDR3 itself, one from Hipparcos, and one from the position difference between Hipparcos and Gaia.  The Hipparcos proper motion is by far the least precise for most stars.  In the current paper, we will only be concerned with the Gaia proper motion and the long-term proper motion given by the position difference between Hipparcos and Gaia.  The difference between these proper motions represents acceleration in an inertial reference frame.  The HGCA received no update for Gaia DR3 \citep{GaiaDR3}; the EDR3 astrometry is identical to the DR3 astrometry for stars with standard five-parameter astrometric fits.

\begin{figure}
    \centering\includegraphics[width=0.9\linewidth]{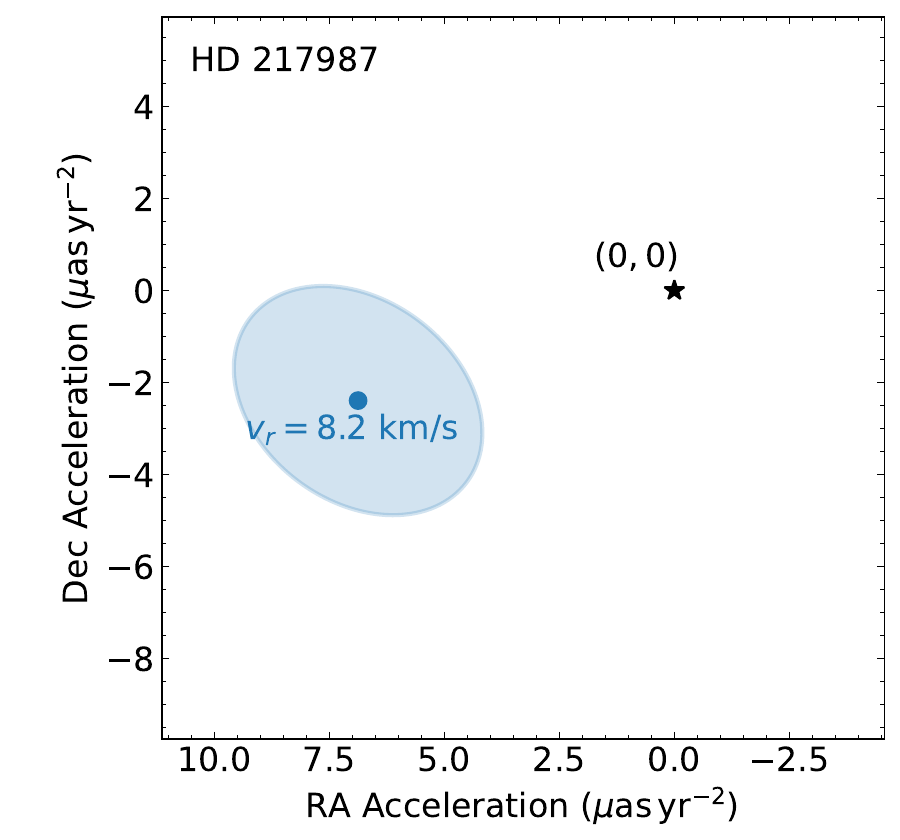}
    \caption{Difference between the long-term and Gaia EDR3 proper motions in the HGCA for HIP\,114046 (=HD\,217987), scaled by their time difference and interpreted as an acceleration.  The error ellipse is computed from the sum of the covariance matrices for the two proper motions.  HIP\,114046 appears to show astrometric acceleration---a nonzero difference in proper motion---at about $2.6\sigma$ significance. \label{fig:firstexample}}
\end{figure}

If deviations from uniform space motion are due only to measurement error, the difference between the Gaia and long-term proper motion should follow a Gaussian distribution described by the sum of their covariance matrices.  We define $\chi^2$ as
\begin{equation}
    \chi^2 = 
    \begin{bmatrix}
        \Delta \mu_{\alpha*} \\
        \Delta \mu_{\delta}
    \end{bmatrix}^T
    \left( {\bf C}_G + {\bf C}_{HG} \right)^{-1} 
    \begin{bmatrix}
        \Delta \mu_{\alpha*} \\
        \Delta \mu_{\delta}
    \end{bmatrix}
    \label{eq:hgca_chisq}
\end{equation}
where $\Delta \mu_{\alpha*}$ and $\Delta \mu_{\delta}$ are the differences between the Gaia and long-term proper motions in R.A.~and Decl., respectively, ${\bf C}_G$ and ${\bf C}_{HG}$ refer to the Gaia and long-term covariance matrices, and $^{-1}$ indicates the matrix inverse.  The use of $\alpha*$ for R.A.~indicates that the $\cos\delta$ factor has been included.  In the absence of physical accelerations, $\chi^2$ as computed from Equation \eqref{eq:hgca_chisq} should follow a $\chi^2$ distribution with two degrees of freedom.  The star-by-star values of $\chi^2$ are included in the published HGCA.  Figure \ref{fig:firstexample} shows an example error ellipse of the proper motion difference between the long-term and Gaia EDR3 proper motions, scaled by one-half the time difference between Hipparcos and Gaia,
\begin{equation}
    \dot{\mu}_{\alpha*} = 2\frac{\mu_{\alpha*,G} - \mu_{\alpha*,HG}}{t_{\alpha*,G} - t_{\alpha*,H}} 
\end{equation}
and
\begin{equation}
    \dot{\mu}_{\delta} = 2\frac{\mu_{\delta,G} - \mu_{\delta,HG}}{t_{\delta,G} - t_{\delta,H}} .
\end{equation}
This star, HIP\,114046 (=HD\,217987), appears to show acceleration at about $2.6\sigma$ significance ($\chi^2 = 9.6$ for two degrees of freedom).

Uniform motion in space results in nonuniform motion when projected onto the celestial sphere.  For most stars, the correction is so small as to be irrelevant at Gaia's precision.  For nearby and/or high proper motion stars, however, the correction can be important.  The HGCA applies this correction by first adopting the Gaia EDR3 parallax, proper motion, and RV to derive a position and velocity in space.  This space velocity is then propagated back in time by the star-by-star differences between the characteristic epochs in R.A.~and Decl.~for Hipparcos and Gaia.  This gives one position difference in R.A.~and one in Decl.~that may be compared with the position differences that would be obtained by multiplying the Gaia EDR3 proper motions by the same time baselines.  The differences between the computed and extrapolated position differences, divided by the time baselines, form the nonlinearity corrections listed in the HGCA and applied to the long-term proper motions.  The nonlinearity corrections applied by the HGCA require the RV.  Where the RV is not measured, it is assumed to be zero.  

The corrections in the HGCA do not account for the finite travel time of light; \cite{Butkevich+Lindegren_2014} derived expressions that do.  For the purposes of the HGCA, the light travel time corrections of \cite{Butkevich+Lindegren_2014} modify the effective time interval between Hipparcos and Gaia, but by an amount that is too small to be important.  Even for Kapteyn's star, which has the largest correction, light travel time effects modify the long-term proper motion by less than its uncertainty.  

\cite{Brandt_2021} showed histograms of proper motion differences in R.A.~and Decl.~to demonstrate the statistical validity of the HGCA calibrations.  They did not show these statistics separately for some of the closest and fastest-moving stars, for which the RV can have a large impact on nonlinear corrections to the astrometric acceleration.   The following sections will explore the impact of the uncertain RV on the apparent astrometric acceleration, the impact on acceleration detections, and the astrometric inference of RV.  

\section{Impact of Radial Velocity on the Astrometric Acceleration} \label{sec:RVimpact}

As discussed in the previous section, a star moving at constant velocity through space will move nonlinearly in projection on the celestial sphere.  The component of this nonlinear motion due to RV is, to first order, 
\begin{equation}
\dot{\bm{\mu}} = -2 \bm{\mu} v_r/D 
\label{eq:rvaccel}
\end{equation}
where $D$ is the distance to the star (\citealt{Schlesinger_1917}; \citetalias{Lindegren+Dravins_2021}).  The use of boldface for the proper motion $\bm{\mu}$ emphasizes that it is a vector, and that the apparent proper motion acceleration $\dot{\bm{\mu}}$ is parallel to the proper motion itself.  

If the stellar RV is uncertain, we can replace $\Delta \mu_{\alpha*}$ and $\Delta \mu_{\delta}$ in Equation \eqref{eq:hgca_chisq} with
\begin{equation}
    \Delta \mu_{\alpha*} - \frac{2\mu_{\alpha*,G}}{D} \left( v_r - v_{r,\rm HGCA}\right) \label{eq:accel_RA}
\end{equation}
and
\begin{equation}
    \Delta \mu_{\delta} - \frac{2\mu_{\delta,G}}{D} \left( v_r - v_{r,\rm HGCA}\right), \label{eq:accel_Dec}
\end{equation}
respectively, where $v_{r,\rm HGCA}$ is the RV used by the HGCA and $\mu_{\alpha*,G}$ and $\mu_{\delta,G}$ are the two components of the Gaia proper motion.  With this substitution, Equation \eqref{eq:hgca_chisq} is no longer a single value for each star, but a quadratic in $v_r$.  It can be written in the form
\begin{equation}
    \chi^2(v_r) = \chi^2_{\rm best} + \left( \frac{v_r - v_{r,\rm best}}{\sigma[v_r]} \right)^2 .
    \label{eq:best_vr}
\end{equation}
In this form, $v_{r,\rm best}$ is the RV that minimizes $\chi^2$, the deviation from an inertial sky path, and $\sigma[v_r]$ may be interpreted as the uncertainty on the RV if the star is believed to be undergoing inertial motion.  Under inertial motion, and assuming uncertainties to be accurately estimated, $\chi^2_{\rm best}$ in Equation \eqref{eq:best_vr} should be distributed as a $\chi^2$ random variable with a single degree of freedom.  This degree of freedom corresponds to the proper motion difference in the direction orthogonal to the measured proper motion.

The result of Equations \eqref{eq:accel_RA} and \eqref{eq:accel_Dec} is that the component of proper motion acceleration parallel to the proper motion must be considered with caution.  In the extreme example where the RV precision is very poor compared with the astrometric precision, the astrometric acceleration measurement becomes effectively one-dimensional.  This is analogous to the case of ground-based astrometry in the presence of differential atmospheric refraction.  In that case, apparent positions in the altitudinal direction depend on elevation, source spectrum, and to a lesser degree, atmospheric temperature and water vapor content.  If an object's spectrum is not known to sufficient precision and/or if an insufficiently characterized atmospheric dispersion corrector was used, astrometric measurements can become corrupted along the altitudinal direction \citep{Chen+Li+Brandt+etal_2022}.  Astrometry in the azimuthal direction, however, is unaffected and retains its native precision.  

Equations \eqref{eq:accel_RA}, \eqref{eq:accel_Dec}, and \eqref{eq:best_vr} show that by changing the adopted RV of a star, we may alter its inferred astrometric acceleration, but only along the direction parallel to its proper motion.  The more precise equations of \cite{Butkevich+Lindegren_2014} accounting for light travel time do not appreciably change this picture.  The effect in their Equation (38) is to change the apparent time interval between two position measurements, which again induces a change in inferred acceleration parallel to the proper motion vector.  Because this effect is small for all stars in our sample, and because it is parallel to the effect of a changing RV, we neglect it.  Accounting for light travel time would very slightly modify the best-fit RVs in Equation \eqref{eq:best_vr}; it is accounted for by the analysis of \citetalias{Lindegren+Dravins_2021}.

\begin{figure}
    \centering\includegraphics[width=0.9\linewidth]{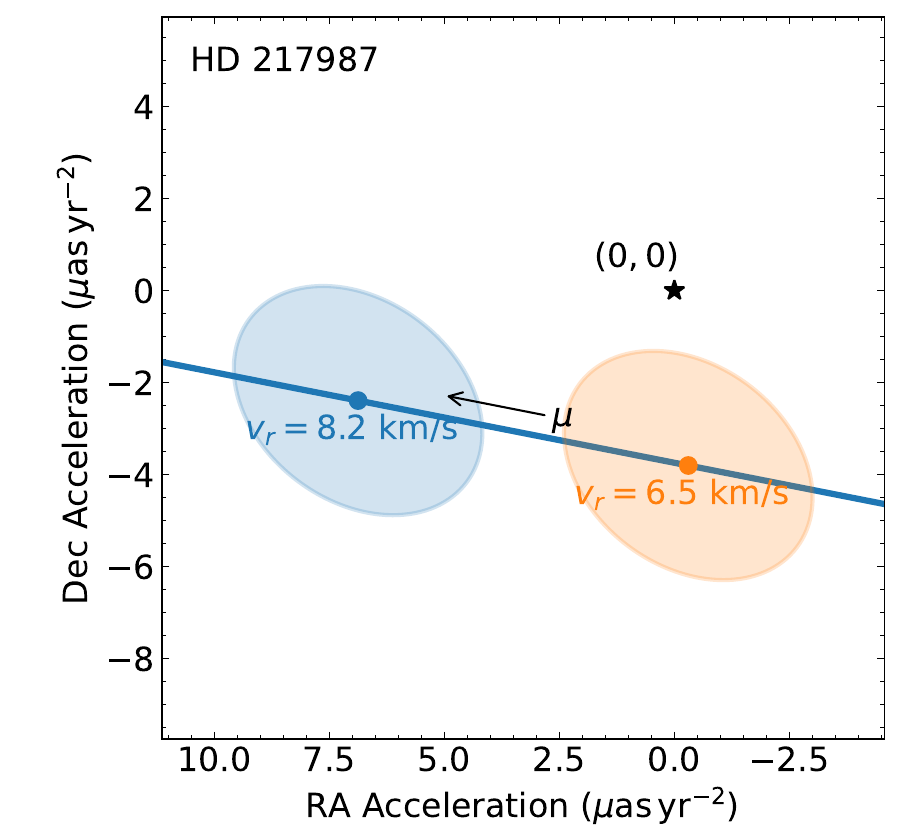}
    \caption{Same as Figure \ref{fig:firstexample}, but with an error ellipse in orange that minimizes the HGCA $\chi^2$ by changing the assumed stellar RV.  This minimum $\chi^2$ is $\approx$2.4, or $\approx$1.5$\sigma$ given the single remaining degree of freedom.  The direction of proper motion is shown with the black arrow and thick blue line. \label{fig:firstexample_withRV}}
\end{figure}

Figure \ref{fig:firstexample_withRV} shows the impact of an uncertain RV on the inferred astrometric acceleration of HIP\,114046, the same star shown in Figure \ref{fig:firstexample}.  The blue line  and black arrow indicate the direction of proper motion.  The orange ellipse represents the error ellipse in the HGCA, but shifted to the RV that minimizes the $\chi^2$ value according to Equation \eqref{eq:best_vr}.  Astrometric acceleration that was significant at nearly $3\sigma$ becomes $\approx$1.5$\sigma$ significant for an appropriate choice of RV.  

\section{Results for Nearby Stars} \label{sec:results_and_notes}

\begin{figure*}
\begin{center}
\includegraphics[width=0.45\textwidth]{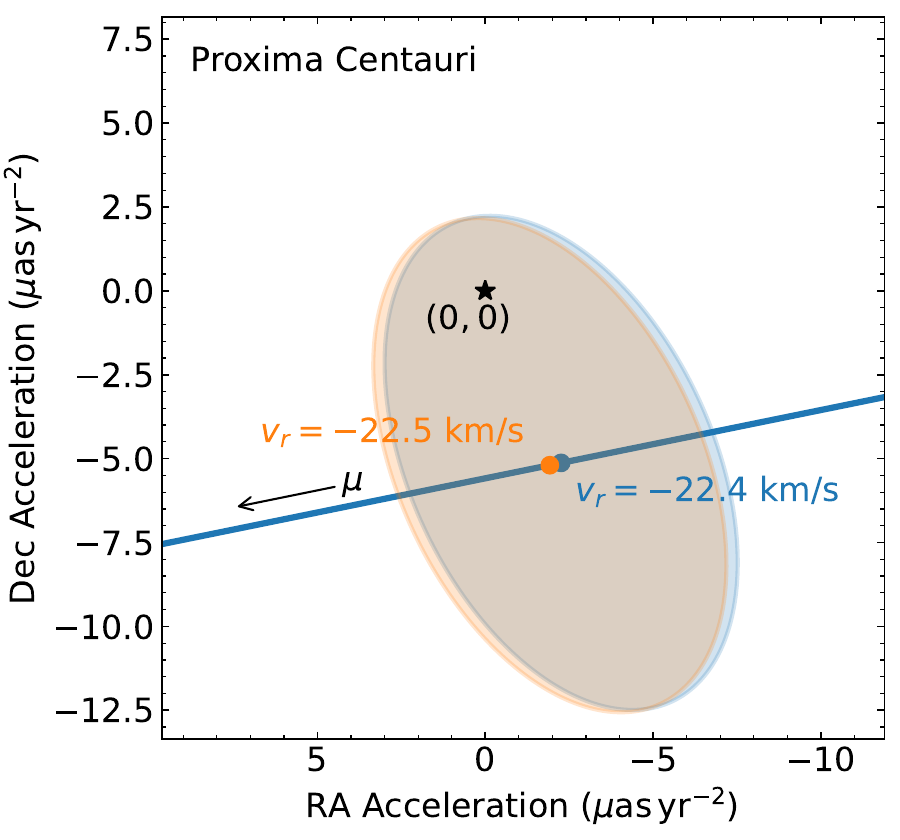}    
\includegraphics[width=0.45\textwidth]{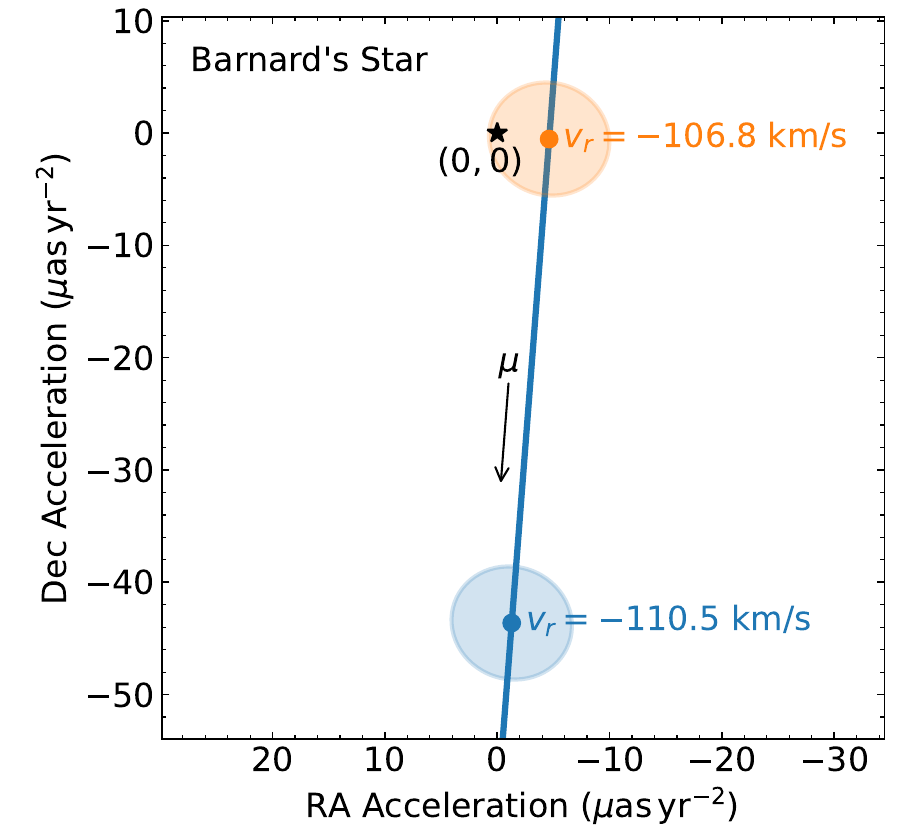}    
\includegraphics[width=0.45\textwidth]{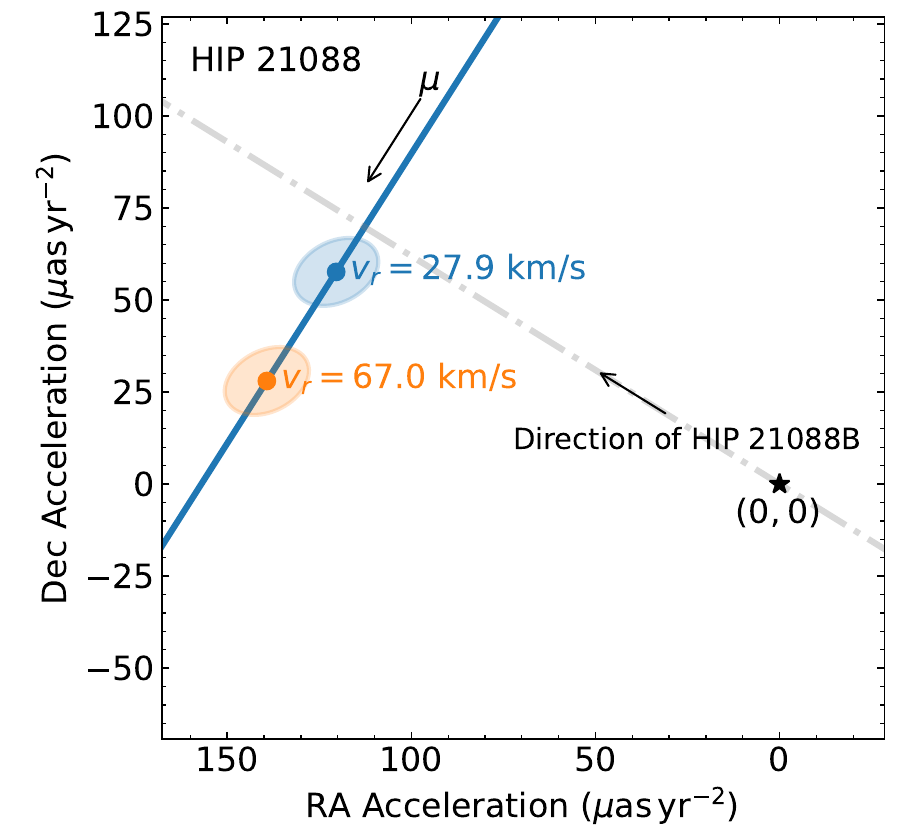}    
\includegraphics[width=0.45\textwidth]{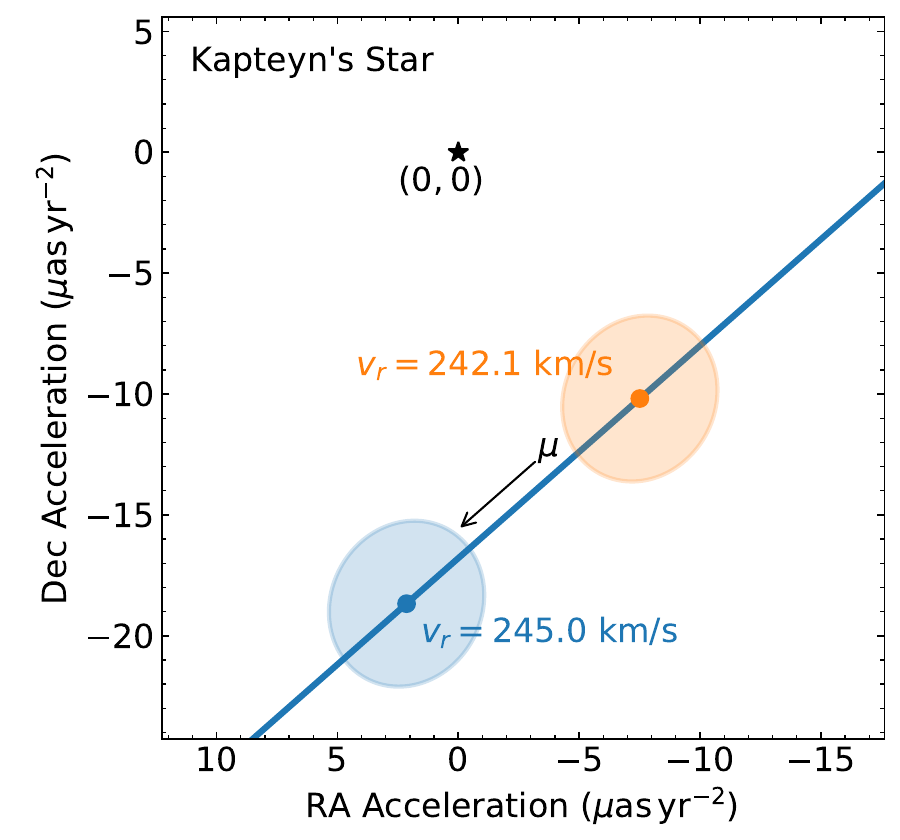}        
\end{center}
\caption{Similar to Figure \ref{fig:firstexample_withRV} but for a selection of nearby stars showing different behaviors.  Clockwise from top left: Proxima Centauri has an astrometric acceleration consistent with zero in the HGCA.  Barnard's Star is a highly significant accelerator in the HGCA, but for a slightly different RV, the significance of this acceleration drops below 1$\sigma$.  Kapteyn's Star, an exceptionally fast mover, remains $\approx$4$\sigma$ inconsistent with constant proper motion even after optimizing the assumed RV.  HIP 21088 (=Stein 2051) remains inconsistent with constant proper motion, but the direction of the measured acceleration matches the orientation of the known white dwarf companion \citep{Giammichele+Bergeron+Dufour+2012,Sahu+Anderson+Casertano+etal_2017}.  \label{fig:astrometricRVplots}}
\end{figure*}

In this section we repeat the calculation shown in Figure \ref{fig:firstexample_withRV} for a selection of nearby stars.  Figure \ref{fig:astrometricRVplots} shows four nearby stars with four representative cases.  Proxima Centauri, the nearest star to Earth, has a $\chi^2$ value in the HGCA of just 0.5 for two degrees of freedom.  Optimizing the RV has little effect on the agreement of the HGCA astrometry from constant proper motion.  Barnard's Star, the star with the highest proper motion, has a $\chi^2$ value of 75 in the HGCA, equivalent to astrometric acceleration at more than 8$\sigma$ significance with the two degrees of freedom.  Optimizing the RV for astrometric agreement yields a $\chi^2$ as low as 0.75, i.e., $<$1$\sigma$ disagreement with constant proper motion.  

Kapteyn's star (HIP 24186), a very fast-moving nearby M dwarf \citep{Kapteyn_1897}, is shown in the lower right panel of Figure \ref{fig:astrometricRVplots}.  It presents a different case.  It is a 5$\sigma$ accelerator in the HGCA ($\chi^2=29.6$ with two degrees of freedom).  After optimizing the stellar RV for inertial motion, Kapteyn's star remains inconsistent with constant proper motion at 4$\sigma$ significance ($\chi^2=16$ with one remaining degree of freedom).  There is currently no clear evidence for companions to Kapteyn's star \citep{Bortle+Fausey+Hallie+etal_2021}.  The astrometric acceleration between Hipparcos and Gaia in the HGCA is $\approx$0.02\,mas\,yr$^{-2}$, or $\approx$37\,cm\,s$^{-1}$\,yr$^{-1}$.  Such an acceleration lies at the edge of what could be detected from the existing $\sim$decade long precision RV monitoring \citep{Anglada-Escude+Arriagada+Tuomi+etal_2014,Bortle+Fausey+Hallie+etal_2021}.  Further astrometric and RV monitoring may clarify the multiplicity status of Kapteyn's star.

Finally, the lower-left panel of Figure \ref{fig:astrometricRVplots} shows the case of HIP 21088 (=Stein 2051).  This star is a $\approx$14$\sigma$ accelerator in the HGCA, with $\chi^2=208$ for two degrees of freedom.  Optimizing the RV for constant proper motion results in a barely improved $\chi^2$ of 200 for one degree of freedom.  However, the direction of acceleration suggested by the HGCA points in the direction of a known white dwarf companion \citep{Giammichele+Bergeron+Dufour+2012,Sahu+Anderson+Casertano+etal_2017}.  This white dwarf is almost certainly the cause of the observed acceleration.

Table \ref{tab:nearbystars} summarizes the results of our calculations for all stars with formal precisions on RV from astrometry alone of better than 12\,km\,s$^{-1}$.  While the $\chi^2$ values for the HGCA represent goodness-of-fit measurements with two degrees of freedom, the $\chi^2$ values in Table \ref{tab:nearbystars} have just one degree of freedom.  The multiplicity column indicates whether the star has known companion(s) with masses and separations that would induce astrometric acceleration at a level comparable to the precision of the HGCA.

\begingroup 
    \setlength{\tabcolsep}{10pt} 
    \renewcommand{\arraystretch}{1.1} 
    \begin{table*}
        \centering
        \caption{Astrometric radial velocities and goodness-of-fit for selected stars}
        \label{tab:nearbystars}
        \begin{tabular}{ c c c c c c}  
HIP ID & RV$_{\rm HGCA}$ (km\,s$^{-1}$) & RV$_{\rm ast}$ (km\,s$^{-1}$) & $\chi^2_{\rm min}$ & multiplicity?\tablenotemark{$a$} & Name or note \\[0.1em] \hline\hline
87937 & $-$110.51 & $-106.79\pm0.43$ & 0.75 & n & Barnard's Star \\
114046 & $+$8.17 & $+6.47\pm0.64$ & 2.4 & n & HD 217987 \\
70890 & $-$22.40 & $-22.46\pm0.76$ & 0.51 & n & Proxima Centauri \\
24186 & $+$244.99 & $+242.13\pm0.77$ & 16 & n & Kapteyn's Star \\
54035 & $-$84.69 & $-82.68\pm0.87$ & 1.8 & y & HD 95735, long-period planet \\
104217 & $-$64.43 & $-84.66\pm0.96$ & 16000 & y & 61 Cyg B \\
439 & $+$25.29 & $+25.3\pm1.2$ & 0.034 & n\\
57939 & $-$97.69 & $-98.5\pm1.4$ & 1.1 & n\\
1475 & $+$11.51 & $-56.7\pm1.5$ & 280 & y & 	Gl 15 A\\
54211 & $+$68.89 & $+78.7\pm2.2$ & 5.2 & y & 	Gl 412 A\\
36208 & $+$18.22 & $+14.3\pm2.3$ & 3.4 & n & Luyten's Star \\
105090 & $+$20.56 & $+19.1\pm2.4$ & 0.94 & n\\
25878 & $+$8.26 & $+18.4\pm3.1$ & 2.1 & n\\
49908 & $-$26.49 & $-25.6\pm3.7$ & 3.8 & n\\
108870 & $-$40.50 & $-71.1\pm3.8$ & 220 & y & $\varepsilon$~Indi, long-period planet \\
104214 & $-$65.94 & $-66.6\pm4.1$ & 320 & y & 61 Cyg A\\
67155 & $+$15.56 & $+24.8\pm4.2$ & 0.073 & n\\
55360 & $+$60.33 & $+64.7\pm4.6$ & 0.58 & n\\
57367 & \ldots & $+29.6\pm4.9$ & 0.67 & n & LAWD 37 \\
85295 & $-$23.70 & $-22.5\pm5.5$ & 4.2 & n\\
80824 & $-$21.61 & $-32.7\pm5.9$ & 0.63 & n\\
19849 & $-$42.62 & $-37.7\pm5.9$ & 9.1 & y & 40 Eri A\\
84478 & $-$0.04 & $+1.4\pm6.2$ & 9.3 & n\\
76074 & $+$15.50 & $+3.0\pm6.3$ & 1.5 & n\\
94761 & $+$35.55 & $+20.8\pm6.6$ & 0.57 & y & 	Gl 752 A \\
91768 & $-$1.07 & $+144.8\pm6.7$ & 250 & y & 	Gl 725 A \\
3829 & $+$263.00 & $-19.8\pm7.0$ & 0.017 & n & Wolf 28, see RV discussion in \citetalias{Lindegren+Dravins_2021} \\
86162 & $-$28.58 & $-16.0\pm7.4$ & 1.1 & n\\
41926 & $+$14.73 & $+16.0\pm7.5$ & 0.42 & n\\
65859 & $+$14.05 & $+38.0\pm7.5$ & 0.013 & n\\
57548 & $-$31.09 & $-30.5\pm7.6$ & 0.14 & n\\
85523 & $-$2.73 & $-23.1\pm7.7$ & 0.16 & n\\
117473 & $-$71.79 & $-67.3\pm7.8$ & 0.064 & n\\
4856 & $+$1.50 & $+11.8\pm8.0$ & 0.23 & n\\
114622 & $-$18.64 & $-10.1\pm8.0$ & 5.2 & y & HD 219134, sub-Jupiter on $\approx$5 year orbit \\
26857 & $+$105.83 & $+122.8\pm8.1$ & 0.0095 & n\\
23311 & $+$21.38 & $+24.7\pm8.2$ & 0.18 & n\\
15510 & $+$87.90 & $+85.9\pm8.3$ & 0.32 & n\\
106440 & $+$12.52 & $-124.7\pm8.4$ & 14 & y & Gl 832, Jovian planet on $\approx$10 year orbit\\
79537 & $+$8.90 & $-3.1\pm8.9$ & 1.1 & n\\
83591 & $+$34.14 & $+38.9\pm8.9$ & 3.3 & n\\
67090 & $+$20.74 & $+22.6\pm9.7$ & 5.0 & n\\
104432 & $-$58.39 & $-61.9\pm9.8$ & 0.44 & n\\
86287 & $-$10.09 & $-0.3\pm9.9$ & 0.0035 & n\\
112460 & $+$0.40 & $-33.1\pm9.9$ & 0.0049 & n\\
58345 & $+$48.33 & $+54\pm10$ & 0.00039 & n\\
86990 & $-$115.00 & $-53\pm10$ & 0.10 & n & Gaia RV is $-43.79$\,km\,s$^{-1}$ \\
74235 & $+$310.88 & $+321\pm10$ & 0.67 & n\\
60559 & $+$50.57 & $+46\pm10$ & 0.027 & n\\
10279 & $-$2.73 & $-8\pm10$ & 0.64 & n\\
73184 & $+$26.81 & $+211\pm11$ & 51 & y & GJ 570 A \\
29295 & $+$4.02 & $-1018\pm11$ & 3700 & y & Gl~229, brown dwarf companions \\
113229 & $+$46.60 & $+62\pm11$ & 0.24 & n\\
10138 & $+$55.23 & $-2821\pm11$ & 8700 & y & Gl~86, white dwarf companion \\
98792 & $-$2.43 & $+23\pm11$ & 0.64 & n\\
86214 & $-$60.00 & $-72\pm11$ & 0.32 & n\\
113020 & $-$1.59 & $+6\pm11$ & 0.12 & n\tablenotemark{$b$} & GJ 876 - see footnote $b$ \\
74234 & $+$311.35 & $+325\pm12$ & 0.00056 & n\\
82588 & $+$44.68 & $+33\pm12$ & 0.68 & n\\
21088 & $+$27.90 & $+67\pm12$ & 200 & y & Stein 2051, white dwarf companion \\
103096 & $-$17.59 & $-26\pm12$ & 0.22 & n\\
            \hline \hline  
        \multicolumn{6}{l}{$a$ - `y' indicates that a known companion is expected to contribute at a level at least comparable to the uncertainty} \\
        \multicolumn{6}{l}{$b$ - The EDR3 Gaia proper motion in the HGCA has been replaced with the proper motion from the DR3 orbital solution} \\
        \end{tabular}  
    \end{table*}
\endgroup

In many cases, our inferred RVs are consistent with those used as inputs for the HGCA.  Most of these stars have best-fit $\chi^2$ values indicating that constant proper motion provides a good fit to the data.  There are a handful of exceptions, like Barnard's Star discussed above.  There are also many stars for which the best $\chi^2$ value is large; nearly all of these have known companions.  The nature of these companions is indicated in the notes, where a designation of ``A'' means that the companion is stellar.  We briefly provide notes on these systems here.  Systems without citations have their companion astrometry taken from Gaia EDR3 \citep{Gaia_EDR3, Lindegren+Klioner+Hernandez+etal_2021}.  

{\it HIP 54035} (=HD 95735): This star has a few-Earth-mass planet on an $\approx$8 year orbit; the RV semiamplitude of the star is slightly more than 1\,m\,s$^{-1}$.  Though this is a small signal, the change in proper motion in the HGCA is just $\approx$0.1\,mas\,yr$^{-1}$.  At a parallax of 393\,mas \citep{Lindegren+Klioner+Hernandez+etal_2021} this converts to 1.2\,m\,s$^{-1}$, i.e., nearly the same as the known RV signal.  HD~95735c shows how, for nearby stars, Gaia's astrometric sensitivity can approach that needed to characterize terrestrial planets.

{\it HIP~104217} (=61 Cyg B): This is the less massive of the 61~Cyg binary; the astrometric acceleration is due to the primary.  This system is analyzed further in the next section.

{\it HIP 1475} (=Gl 15A): The astrometric acceleration is due to the slightly fainter M dwarf Gl~15B; the companion lies $34''$ away in projection.

{\it HIP 54211} (=Gl~412A): The astrometric acceleration is only mildly discrepant from zero, but the companion Gl~412B lies $32''$ away in projection and can account for the observed signal.

{\it HIP 108870} (=$\varepsilon$~Indi): The star hosts a recently-imaged Jovian exoplanet \citep{Matthews+Carter+Pathak+etal_2024}.  This exoplanet is the cause of the observed acceleration, and the acceleration has been used to measure a mass and orbit \citep{Feng+Anglada-Escude+Tuomi+etal_2019}.  

{\it HIP~104214} (=61 Cyg A): This is the more massive of the 61~Cyg binary; the astrometric acceleration is due to the secondary.  

{\it HIP 19849} (=40~Eri~A): While evidence for astrometric acceleration is weak, the 40~Eri~B system (itself a binary) lies $1.\!\!'2$ away in projection.  This would induce an astrometric signal comparable to that observed.

{\it HIP 94761} (=Gl~752~A): While the system is consistent with zero astrometric acceleration, there is a faint companion $1.\!\!'2$ away whose astrometric tug would be comparable to the measurement error.  

{\it HIP 91768} (=Gl~725A): The astrometric acceleration is from the stellar companion Gl~725~B lying 11$''$ away in projection.

{\it HIP 114622} (=HD~219134): The astrometric acceleration signal is weak, but the star hosts a sub-Jupiter exoplanet on a $\approx$5-year orbit \citep{Motalebi+Udry+Gillon+etal_2015}.  The semiamplitude of the RV signal due to the outer planet is $\approx$5\,m\,s$^{-1}$, while the change in proper motion in the HGCA is $\approx$0.15\,mas\,yr$^{-1}$, or $\approx$5\,m\,s$^{-1}$ at the $\approx$6.5\,pc distance to the system.

{\it HIP 106440} (=Gl~832): This star hosts a Jovian planet on a $\approx$10-year period \citep{Bailey+Butler+Tinney+etal_2009}.  The RV semiamplitude of the star due to this planet is $\approx$20\,m\,s$^{-1}$; the proper motion change in the HGCA is $\approx$0.6\,mas\,yr$^{-1}$, or $\approx$15\,m\,s$^{-1}$ at the $\approx$5\,pc distance to the system.  

{\it HIP 73184} (=Gl~570~A): The star hosts a low-mass stellar binary at a projected separation of $\approx$26$''$; these companions are responsible for the observed astrometric acceleration.

{\it HIP 29295} (=Gl~229): This star hosts a well-studied brown dwarf binary on a $\approx$200-year orbit \citep{Nakajima+Oppenheimer+Kulkarni+etal_1995,Xuan+Merand+Thompson+etal_2024,Whitebook+Brandt+Brandt+Martin_2024}.  The brown dwarfs are responsible for the primary star's astrometric acceleration, which has enabled precise mass measurements \citep{brandt_gliese_229b_mass_htof,Brandt+Dupuy+Li+etal_2021}.

{\it HIP 10138} (=Gl~86): The star hosts a close white dwarf companion that is responsible for the observed astrometric acceleration \citep{Mugrauer+Neuhauser_2005}.  The acceleration has been used to measure the white dwarf's mass \citep{Brandt_Dupuy_Bowler_2018}.  

{\it HIP 113020} (=Gl~876): This star has a well-studied planetary system; the most massive planet is a $\approx$2\,$M_{\rm Jup}$ companion on a $\approx$60-day orbit \citep{Marcy+Butler+Vogt+etal_1998,Delfosse+Forveille+Mayor+etal_1998}.  This orbital period is too short to be detectable in the HGCA, but it impacts the Gaia astrometric fit.  We use the barycenter proper motion from the Gaia DR3 two-body orbital fit \citep{Holl+Sozzetti+Sahlmann+etal_2023} instead of the EDR3 proper motion from a five-parameter fit.  

{\it HIP 21088} (=Stein 2051): As noted earlier, the star hosts a white dwarf companion that is responsible for the observed acceleration.

\section{Measurements of Binary Systems} \label{sec:binaries}

In the previous section, we found the RV that minimized the deviation of an observed sky path from constant proper motion.  We now turn to binary systems, in which two acceleration vectors may be measured (one for each star).  This case was also treated by \citetalias{Lindegren+Dravins_2021} from a slightly different perspective than the one we will adopt.  The acceleration vectors for the two stars in a binary should each point toward the opposite star in the system, and they should be antiparallel to one another.  The mass ratio may then be derived from the ratio of the magnitudes of the two acceleration vectors.

If a binary system lacks a well-measured barycenter RV, then the two acceleration vectors may not be exactly antiparallel.  However, in the absence of measurement error and for the correct barycenter RV, they may be made antiparallel.  For the correct mass ratio, the weighted vector sum of the two stars' astrometric accelerations will be zero.  As for the case of single stars, we further assume that the barycenter proper motion $\boldsymbol{\mu}_{\rm sys}$ is known well enough that the direction of perturbations from a nonzero barycenter RV need not be fitted.

For a binary system, then, our model is
\begin{equation}
    {\bf w} = \frac{\boldsymbol{\dot{\mu}}_A + q \boldsymbol{\dot{\mu}}_B}{1 + q} + 2 \frac{\boldsymbol{\mu}_{\rm sys} v_r}{D} \approx 0,
    \label{eq:binary}
\end{equation}
with $v_r$ and $q$ being parameters to be optimized.  An equation for $\chi^2$ may be formed from Equation \eqref{eq:binary}, where the left-hand-side of this equation is multiplied by the inverse of the weighted sums of the covariance matrices of $\boldsymbol{\dot{\mu}}_A$ and $\boldsymbol{\dot{\mu}}_B$:
\begin{equation}
    \chi^2 = {\bf w}^T \left( \frac{{\bf C}_A + q^2{\bf C}_B}{(1 + q)^2} \right)^{-1} {\bf w} . \label{eq:chisq_binary}
\end{equation}
Equation \eqref{eq:binary} has four measurements and two free parameters ($q$ and $v_r$).  However, the factor $q$ multiplies some of the measurements rather than a model.  Assuming no restrictions on the domain of $q$, it is always possible to choose a $q$ that renders the weighted sum of $\boldsymbol{\dot{\mu}}_A$ and $\boldsymbol{\dot{\mu}}_B$ to be parallel (or antiparallel) to $\boldsymbol{\mu}_{\rm sys}$, so the best $\chi^2$ should be zero.  This may not be true if $q$ is bounded, e.g., $0 < q < 1$ on physical grounds.  Thresholds on $\chi^2$ remain meaningful ways to set confidence intervals.  If the acceleration vectors $\boldsymbol{\dot{\mu}}_A$ and $\boldsymbol{\dot{\mu}}_B$ in Equation \eqref{eq:chisq_binary} are parallel to $\boldsymbol{\mu}_{\rm sys}$, $q$ and $v_r$ are fully degenerate.  The vectors $\boldsymbol{\dot{\mu}}_A$ and $\boldsymbol{\dot{\mu}}_B$ only determine $q$ independently of $v_r$ to the extent that they are linearly independent of $\boldsymbol{\mu}_{\rm sys}$.  Equation \eqref{eq:chisq_binary} no longer represents a strictly linear problem because of the form in which the mass ratio $q$ appears, but it may be solved for $v_r$ on a grid of $q$ values.  

We have more information than that represented in Equation \eqref{eq:chisq_binary}: we know that the acceleration vectors should point from each star toward the other.  Denoting the separation unit vector between the two stars as ${\bf \hat{r}}$, the cross product should satisfy
\begin{equation}
    w_{\perp} = \bigg|\left(\boldsymbol{\dot{\mu}} + 2 \frac{\boldsymbol{\mu}_{\rm sys} v_r}{D} \right) \times {\bf \hat{r}}\bigg| \approx 0 \label{eq:wperp}
\end{equation}
for both $\boldsymbol{\dot{\mu}}_A$ and $\boldsymbol{\dot{\mu}}_B$.  We can view this as splitting ${\bf w}$ from Equation \eqref{eq:binary} into three components rather than two: $w_{\perp,A}$,  $w_{\perp,B}$, and  $w_{\parallel}$, where the perpendicular components correspond to Equation \eqref{eq:wperp} for each of Stars A and B and $w_\parallel$ is the component of ${\bf w}$ parallel to ${\bf \hat{r}}$.  To compute $\chi^2$, we must construct the covariance matrix for the set of residuals $\{w_\parallel, w_{\perp,A}, w_{\perp,B}\}$.  We begin by rotating the covariance matrix for each of Stars A and B into the frame defined by ${\bf \hat{r}}$, and denote its components as, e.g.,
\begin{equation}
    {\bf C}_A = \begin{bmatrix}
        C_{\parallel, A} & C_{\Vdash, A} \\
        C_{\Vdash, A} & C_{\perp, A}
    \end{bmatrix}
\end{equation}
where we use $C_{\Vdash, A}$ to denote the off-diagonal term of the covariance matrix.  We have
\begin{equation}
    {\bf C}_w = \begin{bmatrix}
        \frac{1}{(1 + q)^2} \left( C_{\parallel, A} + q^2 C_{\parallel, B} \right) & \frac{1}{1 + q} C_{\Vdash, A} & \frac{q}{1 + q} {C_{\Vdash, B}} \\
        \frac{1}{1 + q} {C_{\Vdash, A}} &  C_{\perp, A} & 0 \\
        \frac{q}{1 + q} {C_{\Vdash, B}} & 0 & C_{\perp, B}
    \end{bmatrix} 
\end{equation}
and
\begin{equation}
    \chi^2 = \begin{bmatrix} w_\parallel & w_{\perp,A} & w_{\perp,B} \end{bmatrix} {\bf C}_w^{-1}
    \begin{bmatrix} w_\parallel \\ w_{\perp,A} \\ w_{\perp,B} \end{bmatrix} . \label{eq:chisq_binary_full}
\end{equation}
With two free parameters, $q$ and $v_r$, the $\chi^2$ value becomes a sum of three terms.  The minimum of Equation \eqref{eq:chisq_binary_full} should be $\chi^2$-distributed with one degree of freedom.  

In practice, we calculate ${\bf \hat{r}}$ by propagating the position of each object in the binary backwards by $\frac{1}{4}$ of the time interval between Hipparcos and Gaia.  This is because the acceleration vector that we use is computed from the difference between the long-term proper motion, with a characteristic epoch halfway between Hipparcos and Gaia, and the Gaia EDR3 proper motion.  

\begin{figure*}
    \includegraphics[width=0.33\textwidth]{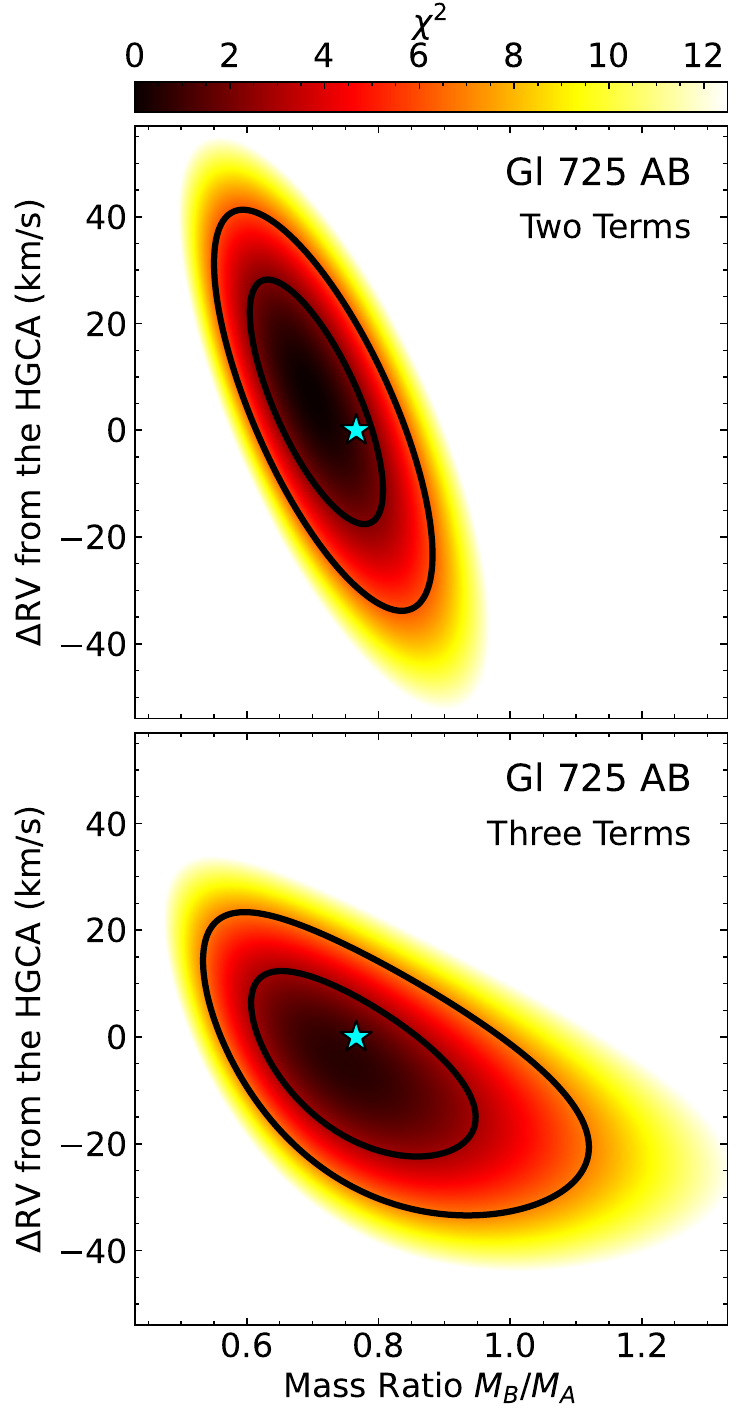} 
    \includegraphics[width=0.33\textwidth]{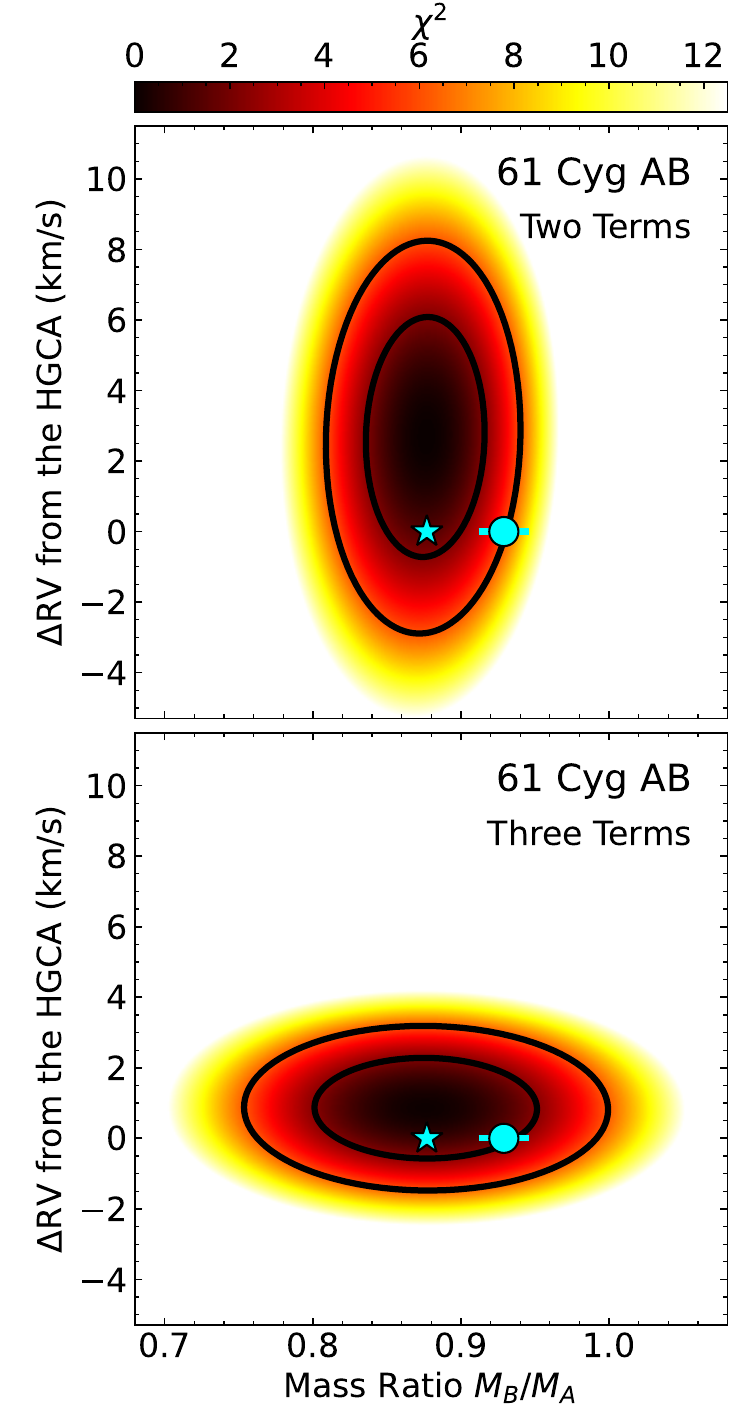}  
    \includegraphics[width=0.33\textwidth]{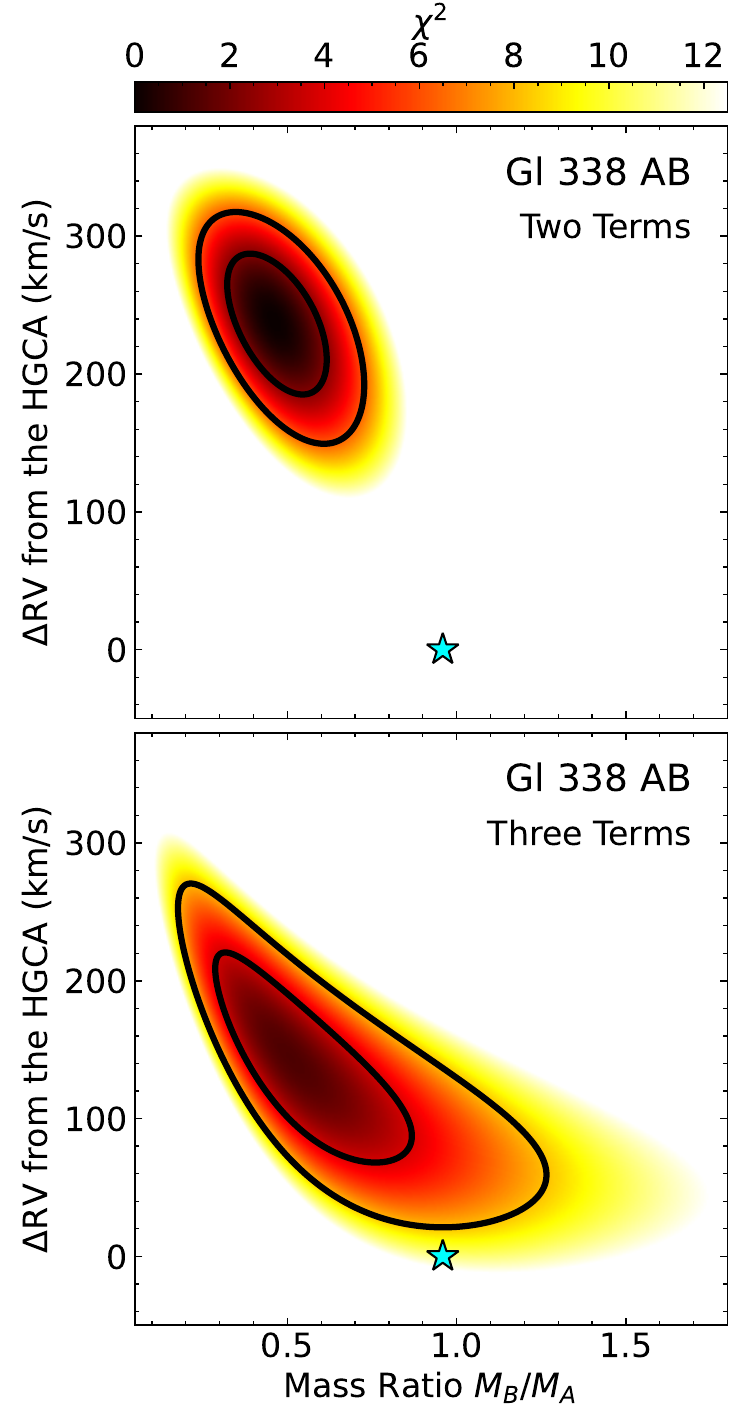}  
    \caption{Left to right: $\chi^2$ for the binaries Gl~725~AB, 61~Cyg~AB, and Gl~338~AB computed using Equations \eqref{eq:chisq_binary} (top) and \eqref{eq:chisq_binary_full} (bottom).  Equation \eqref{eq:chisq_binary_full} can achieve tighter constraints on RV at the partial expense of precision on mass ratio thanks to its separate constraints that the two acceleration vectors in a binary each point to the other star; this is most apparent for the 61~Cyg~AB system (middle panels).  The inner and outer black contours on each panel enclose 68.3\% and 95.4\% of the numerically integrated probability, respectively.  The cyan stars indicate the RVs in the HGCA and the mass ratios inferred from photometry \citep{Kervella+Merand+Pichon+etal_2008,Mann+Feiden+Gaidos+etal_2015,Kervella+Arenou+Mignard+etal_2019}, while the blue circle indicates the dynamical mass measurement of 61~Cyg~AB by \cite{Giovinazzi+Blake+Robertson+etal_2025}.  \label{fig:binary}}
\end{figure*}

Figure \ref{fig:binary} shows $\chi^2$ computed using Equations \eqref{eq:chisq_binary} and \eqref{eq:chisq_binary_full} for the three binary systems in \citetalias{Lindegren+Dravins_2021}.  From left to right, these are Gl~725~AB (HIP~91768 + HIP~91772), 61~Cyg~AB (HIP~104214 + HIP~104217), and Gl~338~AB (HIP~45343 + HIP~120005).  In all cases we also show 1$\sigma$ and 2$\sigma$ confidence intervals computed by integrating $\exp\left(-\chi^2/2\right)$ and determining the threshold enclosing 68.3\% and 95.4\% of the integrated probability.  These differ slightly from the standard $\Delta \chi^2$ thresholds of 2.30 and 6.18 that apply for a Gaussian probability density. 

The probability distributions computed using Equation \eqref{eq:chisq_binary_full} typically show tighter constraints on the RV: Equation \eqref{eq:chisq_binary_full} constrains each star's astrometric acceleration, after correcting for the barycenter RV, to point to the other star in the system.  For both Gl~725~AB and 61~Cyg~AB, and in contrast to \citetalias{Lindegren+Dravins_2021}, we simultaneously recover an RV consistent with the HGCA-adopted value and a mass ratio consistent with infrared photometry \citep{Kervella+Merand+Pichon+etal_2008,Mann+Feiden+Gaidos+etal_2015,Kervella+Arenou+Mignard+etal_2019}.  In the case of 61~Cyg~AB our recovered mass ratio is also consistent with the dynamical mass ratio of $0.929\pm 0.017$ measured by \cite{Giovinazzi+Blake+Robertson+etal_2025}.  There are several additional binaries in the HGCA for which both stars show astrometric acceleration, but all of them have higher-order multiplicity suggested by directly imaged multiples \citep[][and references therein]{Mason+Wycoff+Hartkopf+etal_2001}, non-single-star fits in Gaia DR3 \citep{GaiaDR3}, and/or high renormalized unit weight error (RUWE) values \citep{Lindegren+Klioner+Hernandez+etal_2021}.

For all three systems shown in Figure \ref{fig:binary}, the $\chi^2$ computed using Equation \eqref{eq:chisq_binary_full} admits a broader range of mass ratios than that computed using Equation \eqref{eq:chisq_binary}.  The $\chi^2$ maps are also visibly less Gaussian.  For each of these binaries the mass ratios at zero RV offset from the adopted value in the HGCA are consistent with the mass ratios predicted photometrically.

The binary Gl~338~AB, shown in the right panels of Figure \ref{fig:binary}, shows the largest discrepancy between the mass ratio expected photometrically and the best-fit values given by minimizing Equations \eqref{eq:chisq_binary} and \eqref{eq:chisq_binary_full}.  In \citetalias{Lindegren+Dravins_2021}, the astrometric mass ratio of Gl~338~AB is given as $q=0.452\pm 0.092$.  At 68\% confidence, Equations \eqref{eq:chisq_binary} (top panel of Figure \ref{fig:binary}) and \eqref{eq:chisq_binary_full} (bottom panel of Figure \ref{fig:binary}) give mass ratios  of $0.47^{+0.10}_{-0.09}$ and $0.58^{+0.24}_{-0.17}$, respectively, compared to an expected value of 0.97 \citep{Mann+Feiden+Gaidos+etal_2015,Kervella+Arenou+Mignard+etal_2019}.  

For 61~Cyg~AB and Gl~725~AB, the acceleration vectors of both the primary and the secondary point toward the other component to within measurement error.  For Gl~338~AB, the acceleration vectors of the primary and secondary are both misaligned with the separation vector at $\approx$2$\sigma$: the position angle of the binary at epoch J2010 was $187.\!\!^\circ1$, while the angles of the HGCA $\Delta \boldsymbol{\mu}$ vectors of the A and B components are $178.\!\!^\circ6 \pm 3.\!\!^\circ7$ and $25.\!\!^\circ5 \pm 9.\!\!^\circ1$, respectively.  With the formula for $\chi^2$ given by Equation \eqref{eq:chisq_binary} it is easier for these discrepancies to be absorbed by a large (and implausible) barycenter RV that is offset by several hundred km\,s$^{-1}$ from the spectroscopic RV.  With the formula given by Equation \eqref{eq:chisq_binary_full} this is less true, and the disagreement with the expected RV and mass ratio is less severe.  The discrepancy between the orientation of the separation vector and the acceleration vectors of Gl~338~AB results in a somewhat worse minimum $\chi^2$ value when using Equation \eqref{eq:chisq_binary_full}, $\chi^2_{\rm min} = 1.26$, than for Gl~725~AB ($\chi^2_{\rm min} = 0.54$) and 61~Cyg~AB ($\chi^2_{\rm min} = 0.08$).  

\section{Statistical Properties of the Residuals} \label{sec:statisticalproperties}

We now turn to the statistical properties of the HGCA after optimizing the RVs to minimize deviations from constant proper motion.  This represents a similar test to that shown in the HGCA papers \citep{Brandt_2018,Brandt_2021} where the Gaussianity of the proper motion residuals was demonstrated.  The test here is equivalent to verifying the distribution of the astrometric residuals not in R.A.~or Decl., but parallel and perpendicular to the proper motion.  

\begin{figure*}
\begin{center}
    \includegraphics[width=0.45\textwidth]{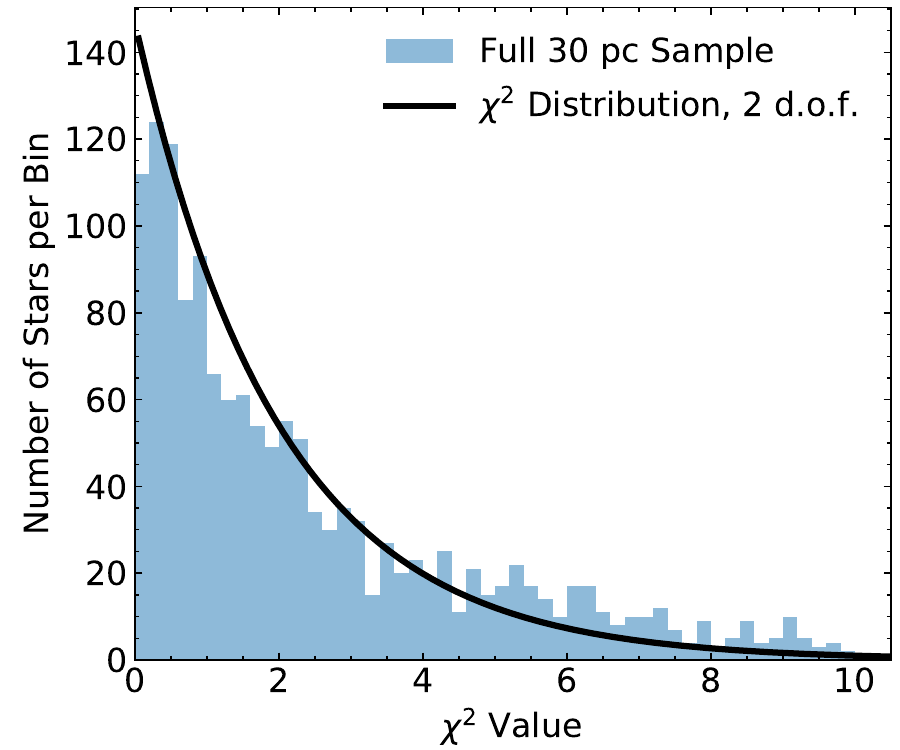}
    \includegraphics[width=0.45\textwidth]{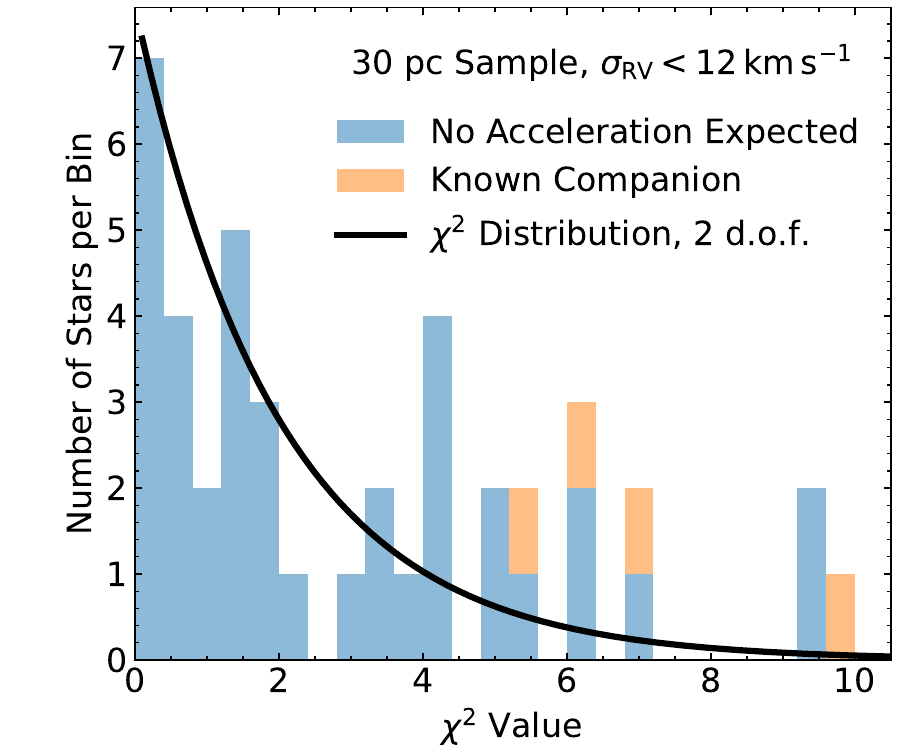}    
\end{center}
    \caption{Left: histogram of $\chi^2$ values from the HGCA \citep{Brandt_2021} for stars within 30\,pc, compared with the $\chi^2$ distribution with two degrees of freedom.  There is a small excess of stars showing apparently real accelerations.  Right: the same $\chi^2$ values, but for a sample of nearby, high proper motion stars whose astrometric corrections are especially sensitive to the assumed RV as described in Section \ref{sec:RVimpact}.  A larger fraction of these appear to be accelerating.  The orange histogram indicates those stars with known companions that are expected to induce significant astrometric accelerations, as indicated by `y' in Table \ref{tab:nearbystars}.  The $\chi^2$ distribution on the right is normalized to the number of stars in the blue histogram, i.e., those without known and astrometrically significant companions. \label{fig:2d_chisqdists}}
\end{figure*}

We begin by comparing the $\chi^2$ distributions for the sample of stars within 30\,pc of Earth without changing the RVs adopted by the HGCA.  In the absence of real accelerators, these should match the $\chi^2$ distribution with two degrees of freedom.  Figure \ref{fig:2d_chisqdists} shows the results.  The left panel shows the full sample of 2589 stars within 30\,pc.  There is a slight excess visible at $\chi^2 \gtrsim 4$; these represent stars undergoing astrometric acceleration.  The right panel of Figure \ref{fig:2d_chisqdists} shows the much smaller sample listed in Table \ref{tab:nearbystars}.  Stars with the multiplicity column in Table \ref{tab:nearbystars} marked by `y' are drawn with an orange histogram.  Even excluding these objects, there is a much higher proportion of apparent accelerators among the nearest stars.  While some of this may be real due to astrometry's inherently better sensitivity to companions around nearby stars, it emphasizes the importance of the use of an assumed RV to correct for perspective acceleration.

We next show the impact of perspective acceleration in coordinates defined by the proper motion unit vector.  We compute the angle between the HGCA proper motion difference and the Gaia proper motion vector; we then rotate both the HGCA acceleration vector and the covariance matrix by this angle.  Finally, we compute the $z$-score along each of these two rotated axes: parallel and perpendicular to the proper motion.  We make no distinction between positive and negative $z$-scores.  The square of the $z$-score in the perpendicular direction matches the minimum $\chi^2$ value listed in Table \ref{tab:nearbystars}, and represents an alternative way of computing this quantity.

\begin{figure*}
    \centering\includegraphics[width=0.9\textwidth]{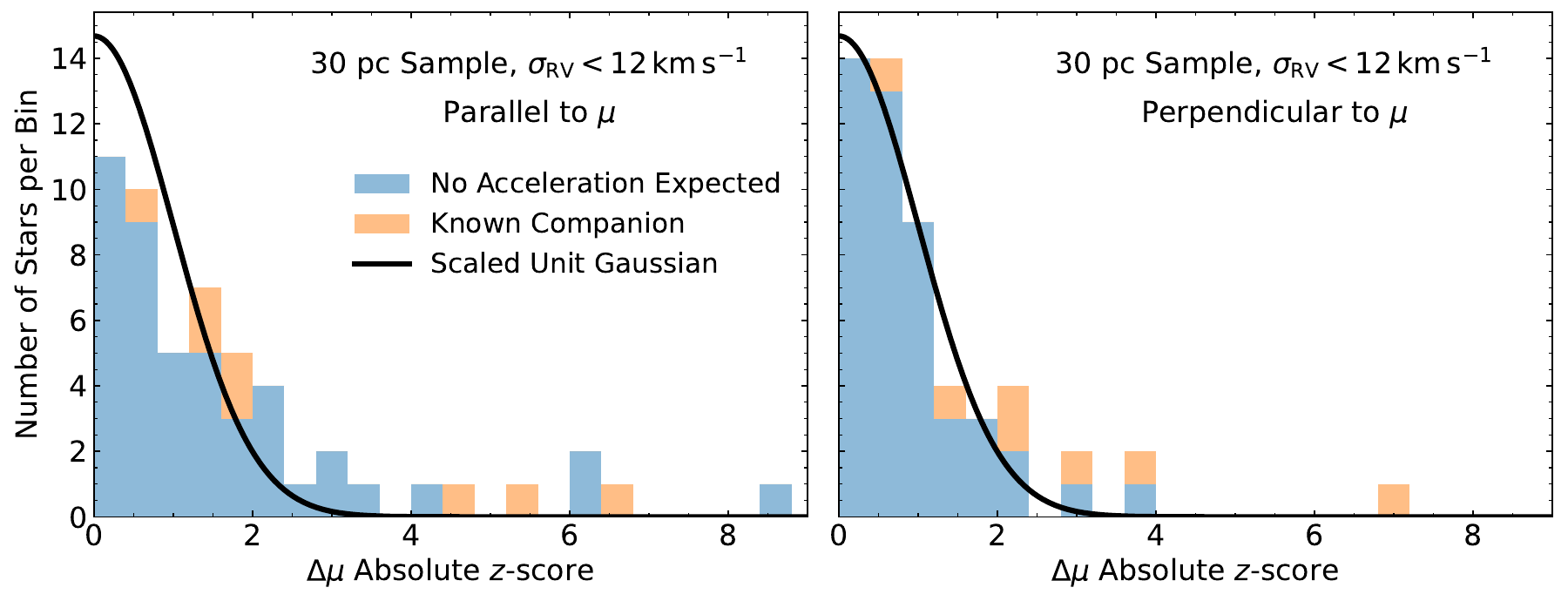}
    \caption{Projections of the HGCA $\delta\mu$ $z$-scores along the directions parallel (left) and perpendicular (right) to the proper motion for the sample of nearby stars listed in Table \ref{tab:nearbystars}.  The unit Gaussian is scaled to the number of sources in the blue histogram, with those with no measurable acceleration expected from any known companions.  The parallel $z$-score is affected by perspective acceleration and deviates significantly from the scaled unit Gaussian.  \label{fig:pdists}}
\end{figure*}

Figure \ref{fig:pdists} shows the $z$-scores of the projected proper motion differences for the sample of stars in Table \ref{tab:nearbystars}.  The left panel shows the one-dimensional $z$-score of the HGCA acceleration along the direction parallel to the Gaia DR3 proper motion; the right panel shows the corresponding $z$-score in the perpendicular direction.  As in the right panel of Figure \ref{fig:2d_chisqdists}, stars marked as multiples in Table \ref{tab:nearbystars} are indicated by orange histograms, while those indicated as apparently single are plotted as blue histograms.  The black curves show unit Gaussians restricted to nonnegative $z$-scores and normalized to the number of stars in the blue histograms.  

The blue histogram in the left panel of Figure \ref{fig:pdists} shows significant deviations from the expected unit Gaussian even for stars where no acceleration is expected.  This deviation from the expected behavior is much reduced in the right panel.  Figure \ref{fig:pdists} provides evidence that, for the nearest and fastest-moving stars, the measured astrometric acceleration parallel to the proper motion must already be treated with caution.  The right panel shows that the measured astrometric acceleration in the orthogonal direction has the expected statistical behavior: the calibrations of the HGCA appear to be effective even for the stars in Table \ref{tab:nearbystars}.  The largest apparently single outlier, with a $z$-score of 4, is Kapteyn's Star and was discussed in Section \ref{sec:results_and_notes} and Figure \ref{fig:astrometricRVplots}.

\section{Impact on Astrometry for Nearby Stars} \label{sec:future_impact}

Figure \ref{fig:pdists} shows that, for nearby stars where perspective acceleration is important, the component of the astrometric acceleration perpendicular to the proper motion is significantly better-behaved statistically than the component parallel to the proper motion.  This suggests that, for such nearby, high-proper-motion stars, the proper motion acceleration vector may be usefully rotated into a basis parallel and perpendicular to the proper motion unit vector.  The impact of perspective acceleration in the direction parallel to the proper motion echoes the impact of differential atmospheric refraction on Earth.  This effect causes a positional shift in a star's apparent position.  The shift depends on the stellar spectrum, the airmass, and the atmospheric conditions, but it only applies in the altitudinal direction.  It is possible to make an analysis robust to the effects of differential refraction by only using measurements in the perpendicular, azimuthal direction.  This principle was used by \cite{Chen+Li+Brandt+etal_2022} to derive a precise mass and orbit of the $\varepsilon$~Indi~BC brown dwarf binary in the presence of strong systematics arising from the use of an insufficiently accurate atmospheric dispersion corrector.

As astrometric sensitivity improves towards terrestrial and ice-giant planets on long period orbits, the importance of accounting for perspective acceleration will grow.  It may be impractical to measure an absolute RV with enough precision to compute a perspective acceleration good enough to take full advantage of the astrometry.  In this case, the largest possible impact on sensitivity to astrometric acceleration would be the loss of one of two orthogonal directions.  Future analyses could separate out these two components as suggested by Figure \ref{fig:pdists} and treat them separately, or treat the component parallel to the proper motion with significantly more care.

At first glance, the loss of one of the two directions of proper motion removes a phase space measurement and makes it much more difficult to measure orbits and masses.  For example, Equations (4)-(7) of \cite{Brandt_Dupuy_Bowler_2018} may no longer be directly applied.  However, if the relative position of the companion responsible for an astrometric tug is known from direct imaging, we can use the fact that the astrometric acceleration should be parallel to this separation vector.  We simply correct our projection of the astrometric acceleration to its full value and accept a somewhat reduced precision, together with the inability to directly check whether the astrometric acceleration and projected separation vectors are parallel.  

\begin{figure}
    \centering
    \includegraphics[width=\linewidth]{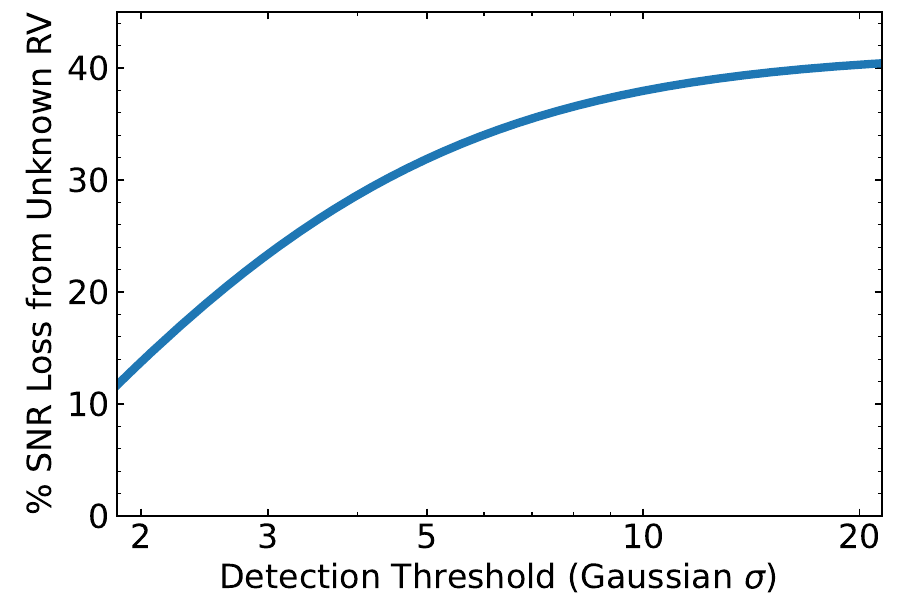}
    \caption{Fractional loss in sensitivity to astrometric acceleration due to the assumed loss of one of two orthogonal directions of the astrometric measurement.  At high detection thresholds, the sensitivity loss asymptotes to $\sqrt{2}-1$, or about 41\%.  }
    \label{fig:sensloss_unknown_RV}
\end{figure}

The worst case for astrometric measurements of nearby stars in light of an uncertain stellar RV would be a total loss of sensitivity in the direction parallel to the proper motion.  Figure \ref{fig:sensloss_unknown_RV} quantifies the loss of sensitivity due to a hypothetical need to use only measurements orthogonal to the direction of proper motion.  The sensitivity loss, while significant, is far from catastrophic.  In the worst case it represents an average reduction in sensitivity by a factor of $\sqrt{2}$.  The impact is smaller at lower detection thresholds, where the loss of one direction also meaningfully reduces noise on the measurement.

The fact that perspective acceleration impacts astrometric acceleration measurements only along the direction of proper motion also suggests that it may be mitigated with a careful choice of observing strategy.  Astrometric missions often have highly asymmetric uncertainties.  Hipparcos and Gaia both varied their scan angles to obtain precise measurements at many orientations.  For a mission targeting individual stars, its sensitivity could be optimized along the direction perpendicular to the (approximately known) proper motion.

Future astrometric missions like TOLIMAN \citep{Tuthill+Bendek+Guyon+etal_2018} and LIFE \citep{Quanz+Ottiger+Fontanet+etal_2022} aim to achieve microarcsecond-sensitivity to nearby stars.  The stellar RV will not affect the detectability of planets on orbital periods shorter than the mission life, but it would affect the detectability of planets on longer periods.  A gas giant or an ice giant on a Saturn-like or wider orbit might only appear as an astrometric acceleration.  In such a case, the stellar RV may need to be accounted for by treating the two components of the astrometric acceleration separately.  

\section{Conclusions} \label{sec:conclusions}

This paper has explored the interplay of stellar RV and astrometric acceleration measurements in the HGCA, with a particular focus on the decomposition of astrometric acceleration measurements into components parallel and perpendicular to the proper motion.  The stellar RV impacts the measured astrometric acceleration parallel to the proper motion, which both limits its utility in searching for real astrometric acceleration and renders it capable of inferring an RV via astrometry.  The component of the astrometric acceleration perpendicular to the proper motion is robust to the effects of an uncertain stellar RV.

For the very nearest and fastest-moving stars in the HGCA, the impact of an uncertain stellar RV can be significant.  We have shown that the component of the astrometric acceleration perpendicular to the proper motion shows much better statistical behavior than the component parallel to the proper motion.  This suggests that, for nearby and fast-moving stars, a user may already consider treating the two components of the astrometric acceleration differently.

The fact that the stellar RV impacts the apparent astrometric acceleration only parallel to the proper motion will be increasingly important for future, high-precision astrometric missions like TOLIMAN \citep{Tuthill+Bendek+Guyon+etal_2018} and LIFE \citep{Quanz+Ottiger+Fontanet+etal_2022} that plan to target very bright and nearby stars.  This fact may render it impractical to detect the tug of long-period companions along one direction, but will leave the measurement along the orthogonal direction unaffected.  The resulting loss in sensitivity, though modest, will depend on the observing strategy and the configuration of an exoplanetary system.  Because it depends on the known stellar proper motion direction, this sensitivity loss should be accounted for in future mission planning.

\noindent {\it Software}: scipy \citep{2020SciPy-NMeth},
          numpy \citep{numpy1, numpy2},
          matplotlib \citep{matplotlib},
          Jupyter (\url{https://jupyter.org/}).

\acknowledgements{I thank Lennart Lindegren for helpful comments on a draft of this manuscript.}
          
\bibliography{refs}{}
\bibliographystyle{aasjournal}

\end{document}